\documentclass[12pt]{article}

 \usepackage{amsfonts}
 \usepackage{amssymb,amsmath}
 \usepackage{graphicx} \usepackage{epstopdf}
 \newcommand{\crlb}[1]{\label{#1}\\[2pt]}
 
 \newcommand{\crld}[1]{\label{#1}}
 \newcommand{\eela}[1]{\quad\hbox{\scriptsize{#1}}\label{#1}\end{eqnarray}}
 \newcommand{\eelb}[1]{\label{#1}\end{eqnarray}}
 
 \newcommand{\newsecb}[2]{\section{#1}\label{#2}\setcounter{equation}{0}}

 \newcommand{\nolabels} {\def\eel{\eelb}\def\eeql{\eeqlb}  \def\crl{\crlb} 
 \def\newsecl{\newsecb}\def\bibiteml{\bibitem} \def\citel{\cite}\def\labell{\crld}}
\newcommand{\eeqla}[1]{\quad\hbox{\scriptsize{#1}}\label{#1}\end{aligned}\end{equation}}
\newcommand{\eeqlb}[1]{\label{#1}\end{aligned}\end{equation}}

\newcommand\publishversion  {\nolabels\setlength{\textheight}{8.3in}\setlength
    {\oddsidemargin}{0in} \setlength{\textwidth}{6.3in}\setlength{\topmargin}{-0.2in}}
\def\beq{\begin{equation}\begin{aligned}}		\def\eeq{\end{aligned}\end{equation}}
\def\be{\begin{eqnarray}}  					\def\ee{\end{eqnarray}}		
   \def\bi#1{\begin{itemize}\item[#1]} 	      	   \def\ei{\end{itemize}} 
   \def\eqn#1{(\ref{#1})}
   	 \def\fn{\footnote}	\def\nm{\nonumber} 

		 \def\a{\alpha}   \def\b{\beta}   \def\d{\delta}   \def\k{\kappa}       

    		  	\def\D{\Delta}    
	    		        		     \def\vv{\varphi}     
 	 		\def\s{\sigma}     	      	 
	     		          		              	\def\w{\omega}  
    		  		\def\dd{{\rm d}} 		\def\HH{{\mathcal H}}  
\def\OO{{\mathcal O}} 		     \def\ZZ{\mathbb{Z}}
 	 	   	    
          \def\pa{\partial}			\def\ra{\rightarrow}	
\def\bra{\langle} 		\def\ket{\rangle}

\def\fract#1#2{{\textstyle\frac{#1}{#2}}}	 	 \def\fractje#1#2{{\scriptstyle\frac{#1}{#2}}}	
\def\ffract#1#2{\raise .2 em\hbox{$\scriptstyle#1\,$}\kern-.3em/\kern-.2em\lower .15 em \hbox{$\scriptstyle\,#2$}}
\def\half{\fract12}			\def\halff{\ffract12}		\def\halfje{\fractje12}
\def\tl#1{\tilde{#1}} 
 			
\def\bpmatrix{\begin{pmatrix}} 			\def\epmatrix{\end{pmatrix}}
\def\bmatrix{\begin{matrix}} 			\def\ematrix{\end{matrix}} 
\def\bcenter{\begin{center}}			\def\ecenter{\end{center}}

\def\lowerheightfig#1#2#3{\(\raise-#1\hbox{\includegraphics[height=#2]{#3}}\)}
\def\lowerwidthfig#1#2#3{\(\raise-#1\hbox{\includegraphics[width=#2]{#3}}\)}
\def\widthfig#1#2{\includegraphics[width=#1]{#2}}
		 
\def\intt{{\mathrm{int}}}   \def\sign{\mathrm{sign}} \def\mmax{{\mathrm{\,max}}}
\def\ontt{{\mathrm{ont}}}

\def\weglaten#1{}	
\publishversion % TO THE EDITOR: please use \publishversion, do not switch on \testprintversion

\begin{document}

\begin{titlepage}
 \title{ \LARGE\bf  Deterministic quantum mechanics: \\
 the mathematical equations\fn{Submitted to the Article Collection: ``Towards a Local Realist View of the Quantum Phenomenon", Frontiers Research Topic, ed. A.~Casado et al.}}
\author{Gerard 't~Hooft}
\date{\normalsize Institute for Theoretical Physics \\ Utrecht University  \\[10pt]
 Postbox 80.089 \\ 3508 TB Utrecht, the Netherlands  \\[10pt]
e-mail:  g.thooft@uu.nl \\ internet: 
http://www.staff.science.uu.nl/\~{}hooft101/ }
 \maketitle

\begin{quotation} \noindent {\large\bf Abstract } \\[10pt]
Without wasting time and effort on philosophical justifications and implications, we write down the conditions for the Hamiltonian of a quantum system for rendering it mathematically  equivalent to a deterministic system. These are the equations to be considered. Special attention is given to the notion of `locality'. Various examples are worked out, followed by a systematic procedure to generate classical evolution laws and quantum Hamiltonians that are exactly equivalent. What is new here is that we consider interactions, keeping them as general as we can. The quantum systems found, form a dense set if we limit ourselves to sufficiently low energy states.  The class is discrete, just because the set of deterministic models containing a finite number of classical states, is discrete. In contrast with earlier suspicions, the gravitational force turns out not to be needed  for this; it suffices that the classical system act at a time scale much smaller than the inverse of the maximum scattering energies considered.
\end{quotation}
\end{titlepage}

\setcounter{page}{2}

\def\inn{{\mathrm{in}}}  \def\outt{{\mathrm{out}}} 
\def\Pl{{\mathrm{Planck}}}
\newsecl{Introduction: ontological quantum mechanics}{intro}
Discussions of the interpretation of quantum mechanics\,\cite{Dirac-1930}--\cite{superdet-2020} seem to be confusing and endless. This author prefers to consider the mathematical equations that make the difference. Having the equations will make the discussion a lot more straightforward. Here, we reduce the theory of quantum mechanics to a mathematical language describing structures that may well evolve deterministically. The language itself is equally suitable for any system with classical or quantum evolution laws.\fn{Systems that are irreversible in time can also be described this way but require adaptations not considered in this paper.}

Every state a system can be in is represented by a unit vector. We are interested in  distinguishing ``ontological states". These are unit vectors that are mutually orthogonal and have norm one; they  form an orthonormal basis of  Hilbert space. We can distinguish finite dimensional Hilbert spaces and infinite dimensional ones.  A system is said to be deterministic if ontological states evolve into ontological states.\cite{GtHCA}--\cite{BJV-2000}.

We use Dirac's bra-ket notation\,\cite{Dirac-1930} both for classically evolving systems and for quantum mechanical ones. A state is indicated as \(|n\ket\),  where \(n\) stands short for some description of this state. Often, we simply enumerate all available states, choosing \(n\in\ZZ\), the set of integers. Alternatively, we can have states \(|x\ket\), where \(x\) takes the values of all real numbers, or we can have vectors, \(|\vec x\,\ket,\ |\vec n\,\ket\,, \ \dots\).

The first models we consider will  seem to be too simplistic to represent all interesting and relevant quantum systems in general. These basic models must be looked upon as building blocks for a more complete theory for deterministic quantum mechanics\fn{The words `deterministic' and `ontological', or `ontic' for short, are almost interchangeable in this paper.}. At the end they will be coupled by (deterministic) interaction Hamiltonians. What is produced in this paper is a generic machinery to be employed in these constructions. We do realise that further streamlining will make our fundamental observations more transparant.

Determinism can be recognised by analysing the eigenvalue spectrum of the Hamiltonian\,\cite{GtHCA}. At first sight, it seems that only Hamiltonians that are linear in the momenta can represent ontological systems, but this happens only if one assumes the system to be strictly continuous.  If we assume the time coordinate to be on a (very dense) lattice, the Hamiltonian eigenvalues are periodic, \emph{i.e.}, these eigenvalues can be forced to sit on a finite interval. If temporarily we limit ourselves to a single, isolated, elementary building block of a more general quantum system, allowing for only a finite number of states, we may assume it to be periodic in time. As we shall observe, deterministic periodic systems can be identified with quantum harmonic oscillators; these have quite realistic Hamiltonians complete with one stable ground state. If time is quantised, we find a useful 
internal \(SU(2)\) symmetry. In that case, there is not only a vacuum state (the state with lowest energy), but there is also an `antivacuum', the state with highest energy. Antivacua may play an important role in black hole physics.\,\cite{GtHBH-2016}

Special attention is needed for the concept of locality. For instance, in a free quantum particle in one spacial dimension, with a fairly general expression of the kinetic energy function \(T(p)\), we can define an appropriate ontological operator, its `beable', but it is a non-local function of \(x\) and \(p\). Such models cannot directly be applied to physically realistic scenarios; instead, they are used as intermediate steps towards more satisfactory procedures, as will be explained.

We claim that locally ontological and deterministic systems can be constructed that nevertheless feature quantum mechanical properties, including models as complex as the Standard Model. These deterministic systems take the form of `cellular automata' \,\cite{Ulam-1952}--\cite{Wolfram-2002}. Formally, there is a limit to the accuracy by which this can be done, but if, as is suspected, the scale at which determinism becomes manifest, is the Planck scale, then we shall have an enormous range of ontological theories that can reproduce all known data quite accurately. They will all be fully quantum mechanical in the usual\,\cite{Merzbacher-1961}--\cite{Das-1986} sense.

This includes the Born interpretation of the absolute squares of amplitudes as representing probabilities.\,\cite{ Born-1926}. Here, we are still free to use various different definitions of what `probability' might mean; in all cases, the definition will be passed on to the wave functions being considered. In our case: \emph{probability in \(=\) probability out}: the probability for the outcome of an experiment is directly related to the probability for the initial state that was chosen. The Schr\"odinger equation just passes on that concept of probability from initial to final state.

Second quantisation will be a natural ingredient, a process that restores not only local causality, but also positivity of the Hamiltonian; in principle, it works just as in Dirac's formalism for quantised fields, but in our formalism, the interactions are taken care of in a way that is somewhat more complicated than in the Standard Model. We find that second quantisation serves a double purpose: it restores special relativity without generating negative energy states\,\cite{ItzyksonZuber-1980}--\cite{GtH-ConcBasis-2007}, and it also restores locality without sacrificing ontology. Second quantisation in our formalism has been elucidated in Ref.\,\cite{GtHCA}.

Our paper is set up as follows. Deterministic models may be seen as consisting of elementary cells inside which the data just oscillate in periodic orbits. We first explain such cells in Section \ref{SU2}. We explain and exploit the  \(SU(2)\) symmetry that shows up and comes out handy, because this symmetry is so well-known and studied.

The idea that deterministic systems of this kind can be described \emph{as if} they were quantum mechanical, is briefly illustrated in  section~\ref{wave}. Hilbert space is an extremely useful device, but it should \emph{not} be taken for granted that Hilbert space is prerequisite in elementary quantum theory. In contrast however, the notion that energies, momenta, and even space-time coordinates, are \emph{quantised}, is very essential, and consequences of this are immensely important for our understanding of nature.

We then go into the direction of thinking about particles in an ontological language. This should be possible, but it seems to give a serious clash with the most elementary concepts of locality. A single quantum particle of the kind we frequently encounter  in atomic physics, in solids, and in most of the elementary particles, behaves in a way that does not seem to allow directly for a deterministic interpretation. We do describe what happens in Sections \ref{box} and \ref{kin}. A single, isolated particle can be well described if its kinetic energy is just a linear function of its momentum; if the kinetic energy is anything more general than that, we do have an interesting ontological variable, but it is non-local. This jeopardises any attempt to add some kind of ontological potential term for the particle.

Section \ref{tobe} sets the stage for what comes next, and then comes the most important part of this paper. A reader who wants to go directly into the deep should mainly be interested in Sections~\ref{interact} and \ref{connect}. Here, we join our elementary cells into a construction where they interact, again allowing only deterministic interaction laws. We do things that are normally not considered: allow for evolution laws that directly exchange ontological states. Surprisingly perhaps, this leads to interaction Hamiltonians that are as general and as complicated as what we usually only encounter in genuine quantum systems.
We emphasise that this proves that the distinction between `quantum interactions' and `classical interactions' is artificial, and was the result of our lack of phantasy concerning the interactions that are possible, even if we limit ourselves to what usually is called `deterministic'. As will be seen explicitly in Section~\ref{connect}, what is normally thought of as being `quantum mechanics' can be attributed to the effect of fast, almost hidden, variables.

A picture emerges of quantum mechanics being an auxiliary device, it is a scheme allowing us to perform statistical investigations far beyond the usual procedures in condensed matter physics and thermodynamics. What is found can be referred to as a `deterministic local field theory', which might be able to dethrone  `quantum field theory'.

Our mathematics may hint at what might be the main  fundamental cause of the apparently true `quantum' nature of our world.
The source of the apparent indeterminism in quantum mechanics appears to be \emph{timing}. When two systems, just slightly separated from one another, are allowed to interact, we have to realise that both systems contain internal parts that oscillate at time scales that are very small even according to time standards used in elementary particle physics. The only way to register what happens when they interact, is to project away the ultra fast time components of both systems. This can only be done by selecting sufficiently low energy eigen states of the Hamiltonian, which is a procedure that can only be done by introducing Hilbert space. Today, physicists only have access to the very lowest energy states, and these can only be addressed in quantum mechanical language. We leave it to the philosophers to expand on such observations or suspicions.

\newsecl{The standard quantum mechanical Hamiltonian for continuous systems}{continuous}
Historically, quantum mechanics was first studied for continuous systems, that is, coordinates and momenta are continuously defined on \({\mathbb R}^n\) spaces. The generic Hamiltonian \(H\) is then written as
	\be H=T(\vec p\,)+V(\vec x\,)+\vec A(\vec x\,)\cdot p\ , \eel{contHam}
where \(\vec x\) and \(\vec p\) are the usual, continuously defined, coordinates and momenta, obeying 
	\be [x_i,\,p_j]=i\d_{ij}\ . \ee
the third term is actually the simplest. A Hamiltonian having \emph{only} this term, describes a completely deterministic system, since the Hamilton equations then read:
\be H=\vec A(\vec x\,)\cdot\vec p\ ,\qquad \frac\dd{\dd t}\vec x=\frac{\pa H}{\pa\vec p}=\vec A(\vec x)\ ,\eel{magnHam}	
while the time derivative of \(\vec p\) is not directly needed. We point out that, in some elementary sense, \emph{all} deterministic evolution laws can be cast in the form \eqn{magnHam}, so this is actually a very important case. 

In the usual quantum systems, we have as a central unit the kinetic term \(T(\vec p\,)\). Usually, but not always, it takes the form \(T(\vec p\,)=\half\vec p\,^2\). In that case, the `magnetic' term \(\vec A\cdot\vec p\) plays a more secondary role. This case is already considered `essentially quantum mechanical', displaying the characteristic interference patterns. Still, particles that basically move in straight lines might not be the most interesting physical things, so the third term, \(V(\vec x)\), where the function \(V\) can be almost anything, would be needed to cover almost all systems of physical interest. This is the most difficult case from the present point of view.

In our discussion, we take one important step backwards: space, time, and often also momentum, will be kept discrete. The continuum limit can always be considered at some later stage. The question is, whether we can handle the interesting case of the general Hamiltonian \eqn{contHam}, as the continuum limit of the models that will be discussed now. These models typically describe only a finite-dimensional vector space, for the time being.

\newsecl {The periodic model, and  its \(SU(2)\) symmetry.}{SU2}
Our elementary building block will be a system or device that updates itself at every time step, of duration \(\d t\), and, after a period\  \(T=N\,\d t\),   it returns to its initial position.
The elementary, ontological states are \(|k\ket^\ontt\), where \(k\) is an integer, \(k=0,\dots N-1\).  Note that, at this stage, there is no quantum mechanics in the usual sense, but we shall use quantum mechanics merely as a language.\cite{GtHCA}  The ontological states considered here are closely related to the concept of `coherent states' that have a long history going back to R.J.~Glauber\,\cite{Glauber-1963}.

The evolution operator  over one time step, \(U(\d t)\),  is simply defined by
\be |k\ket_{t+\d t}^\ontt= U(\d t)\,|k\ket_{t}^\ontt=|k+1\!\mod N\ket_{t}^\ontt\ . \qquad U(\d t)= 
 \bpmatrix 0&\cdots&0& 1\\ 
 		1&\cdots&0& 0\\ 
		 \vdots&\ddots&&\vdots\\ 
		0&\cdots& 1&0 \epmatrix \ .
\eel{ontok}
The matrix \(U\) is easily diagonalised by using the finite Fourier transformation:
\be |k\ket^\ontt=\fract 1{\sqrt{N}}\sum_{n=0}^{N-1}e^{-2\pi ink/N}|n\ket^E\ ,\qquad  
|n\ket^E=\fract 1{\sqrt{N}}\sum_{k=0}^{N-1}e^{2\pi ink/N}|k\ket^\ontt\ ,\qquad  \eel{FF}
where \(|n\ket^E\) are the energy eigenstates. We have
\be U(\d t)|n\ket^E=e^{-2\pi in/N}|n\ket^E\ ,\ee
and we can define the Hamiltonian matrix \(H\) by imposing
\be U_{\d t}=e^{-iH\d t}\ ; \qquad H|n\ket^E=\fract {2\pi n}{N\d t}\,|n\ket^E
=n\w\, |n\ket^E\ ;\qquad \w=\fract{2\pi}T  \ee
(Note that the \emph{ground state energy} has been tuned to zero here; we shall also do this when the harmonic oscillator is discussed; the reader may always consider `corrected' definitions where the ground state has energy \(\half\w\)). 

With the Fourier transform \eqn{FF}, one can easily determine how \(H\) acts on the ontological states \(|k\ket^\ontt\).

Our mathematical machinery becomes more powerful when we realise that the energy eigenstates may be regarded as 
the eigenstates \(|m\ket\) of \(L_3\) in a three dimensional rotator. Let the total angular momentum quantum number \(\ell\) be given by
\be N\equiv 2\ell+1\ ; \qquad n=m+\ell\ , \qquad H=\w(L_3+\ell)\ . \ee
Then define the (modified) operators \(p\) and \(x\) by
\be L_1\equiv p\sqrt\ell\ ;\qquad L_2\equiv x\sqrt\ell\ ;\qquad [\,x,\,p\,]=-iL_3/\ell=i(1-H/\w\ell)\ ;\labell{xpellcomm}\\
\ell(\ell+1)=L_1^2+L_2^2+L_3^2\  \ \ra\ \ 
H=\fract \w{1-H/2\w\ell}\,\half(p^2+x^2-1)\ , \eel{Hell}
and we see that, when \(\ell\) tends to be large  (while the energy \(H\) and the fundamental time step \(\d t\) are kept small), this reduces to the standard Schr\"odinger equation for the harmonic oscillator (the ground state value \(\half\w\) has been subtracted from this Hamiltonian).
Both \(x\) and \(p\) take \(2\ell+1\) values, and they span the entire Hilbert space, but they are not ontological. We identify the original states \(|k\ket^\ontt\), Eq.~\eqn{ontok}, as our ontological states.

The operators \(L_\pm=L_1\pm i L_2\) play the role of creation and annihilation operators: they add or subtract one unit \(\w\) to the energy of a state. However, in the upper half of the spectrum, \(L_\pm\) interchange their positions: the algebra is such that \(L_+\) can no longer add energy above the limit \(m=+\ell\), so that the energy spectrum stretches over the finite interval \([0,\,2\ell \w]\). There is an obvious symmetry \(H\leftrightarrow 2\ell\w-H\). And therefore,\begin{quote}
\indent \emph{the vacuum state \(|0\ket^E\) has a counterpart, the ``antivacuum"  \(|\,2\ell\,\ket^E\), where the energy is maximal.}\end{quote}
Not only in second quantisation, but also in black hole physics, such states play an important role. This is an inevitable consequence of our desire to find a finite dimensional theory for quantum mechanics.

Both \(x\) and \(p\) are discrete operators, just like the Hamiltonian:
\be H=\w n\,,\quad x=r/\sqrt \ell\ ,\quad\ p=s/\sqrt\ell,\qquad r,s\ \hbox{ are integers }
\in [\,-\ell, \,\ell\,] \ , \ee
but, due to the modification \eqn{xpellcomm} in their commutation rule, the unitary operator linking the eigen states \(|r\ket^x\) and \(|s\ket^p\) is more complicated than usual. This happens as long as the elementary time step \(\d t\) is kept finite. In the limit \(\d t\ra 0\) we recover the usual harmonic oscillator. The ontological states then run continuously along a circle.  The unitary operator linking the \(p\) basis to the \(x\) basis for finite \(\ell\)  can be obtained from the matrix elements of the operator \(e^{\halfje i\pi L_1}\) in the basis where \(L_3\) is diagonal. Going to the conventional notation of \((\ell,m)\) states, one can show that
\be{}^x\bra r|s\ket^p=\bra m|e^{\halfje i\pi L_1}|m'\ket\ ,\quad \hbox{where}\quad m=r,\ \ m'=s\ ; \eel{matrix31}
this is derived by first noting that \(L_1\) is diagonal in a basis that is rotated by \(90^\circ\) in angular momentum space, compared to the basis where \(L_2\) is diagonal, and then interchanging \(L_1, \ L_2\) and \(L_3\). The matrix elements \({}^x\bra r|s\ket^p\) can be deduced from recursion relations\,\fn{There is some freedom of phase factors in the definition of the states \(|r\ket^x\) and \(|s\ket^p\).}  such as
\be 2r{\,}^x\bra r|s\ket^p=\sqrt{(\ell+1-s)(\ell+s)}\,{\,}^x\bra r|s-1\ket^p+\sqrt{(\ell+1+s)(\ell-s)}\,{\,}^x\bra r|s+1\ket^p\ ,\qquad \eel{recursion}
in combination with \({\,}^x\bra r|s\ket^p={\,}^p\bra s|r\ket^{x\,*}\) and more. The result can be pictured as in Fig.~\ref{matrixfig}.
\begin{figure}\bcenter\widthfig{200pt}{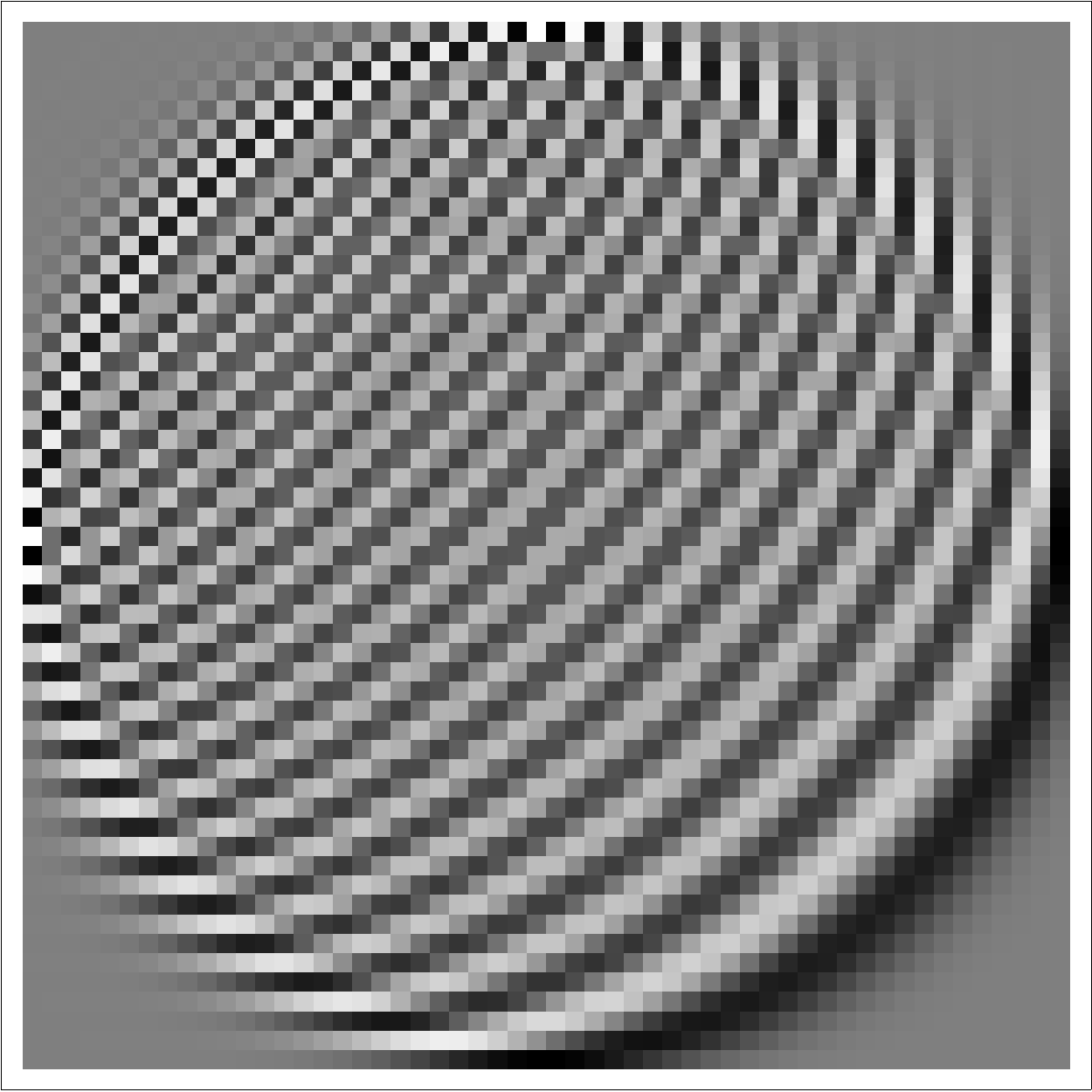}\ecenter\caption{\footnotesize 
The matrix \eqn{matrix31}, for the case \(\ell=53/2\).  
The values of a real solution for the matrix elements are shown. Black is maximal, grey \(\approx 0\), white is minimal.
Notice that, these elements quickly converge to zero outside the circle \(m^2+m'\,^2=\ell\,^2\) (obtained using Mathematica\(^\copyright\) to integrate Eq.~\eqn{recursion}). \labell{matrixfig}}\end{figure}Å

We can define beables \(S_\pm\) as
\be S_\pm|k\ket^\ontt=e^{\mp\, 2\pi i k/N}|k\ket^\ontt\ , \qquad S_\pm|n\ket^E = |n\pm 1\mod N\ket^E\ ,\ee
see Eq.~\eqn{FF}. They are related to \(L_\pm\) as follows:
\be L_+=\sqrt{(n+1)(2\ell-n)}\, S_+\ ,\quad L_-=\sqrt{n(2\ell+1-n)}\,S_-\ . \eel{Spm}
We see here that, due to the factors inside the square roots, the quantum numbers \(n\) are now limited both from below (\(n\ge 0\)) and from above (\(n\le 2\ell\)), but these same square roots imply that the numbers \(k\) for the ontological states do not represent the eigen states of any of the angular momentum operators \(L_i\), they are superimposed to form such eigen states,
so that \(L_\pm\) aren't beables, and neither are \(L_1\) and \(L_2\), or \(p\) and \(x\), Eq.~\eqn{xpellcomm}.

In the limit \(\ell\ra\infty\), the second factors in the square roots \eqn{Spm} become constants, so that, indeed, \(L_\pm\) act as creation operator \(a^\dag\) and annihilation operator \(a\).
The square root of \(H\) that one may recognise in Eqs.~\eqn{Spm}, relating beables with the more familiar \(x\) and \(p\) coordinates, will be encountered again in Section~\ref{kin}.

We elaborated on these mathematical rules in this Section, just because angular momenta are so familiar, and also to emphasise that the finite periodic model (the system with both \(\d t\) and the period \(T\) finite) can be examined using this well-known algebra.

The limit \(\d t\ra 0\) turns this system into a point moving continuously along a circle, which in every respect behaves just like the standard harmonic oscillator, as we shall see in section~\ref{tobe}, but this limit must be taken with some care.

\newsecl{On the wave function generated by a periodic ontological system}{wave}
The periodic ontological system is characterised by a classical kinetic variable defined on a finite interval with periodic boundary conditions. No generality is lost if we assume this to be an angle \(\vv\)  defined on the period \([\,0,\,2\pi\,]\), implying
\(\vv(t)=2\pi k/N\). In the continuum limit, we define the evolution to be  \(\dd\vv(t)/\dd t=\w\), with \(\w\) fixed.

To understand what the quantum wave function here means, we have to assume time to be sliced in small and equal time steps \(\d t=2\pi/N \w\), where \(N\) is a large integer. In the previous section, it was found that the energy \(E\) must then lie in an interval of length \(\w N\). We do have the freedom to define the phase factors of the energy eigen states, and those of the ontological states \(|\vv\ket\), such that this interval is exactly \([0,\,\w N]\). The importance of this choice is that an energy eigen state with energy \(n\w\) evolves as
\be \psi_n\ \propto\ e^{-i\w n t}\ \propto\ \big(e^{-i\vv(t)}\big)^n\ , \ee
and writing \(z\equiv e^{-i\vv}\) one finds that all energy eigen states are positive powers of \(z\). Any wave function expanded in these energy eigen modes is therefore regular for all \(z\) that lie inside the unit circle. To arrive at this insight it was crucial that we start with a discrete lattice in the time variable, where the time spacings \(\d t\) may be chosen arbitrarily small but not zero.

Our motivation for writing this short section was to demonstrate how the elementary building blocks for deterministic models generate the basis elements of complex-valued wave functions in Hilbert space.\fn{It might be interesting that in the ``classical like rewriting" of standard QM by Heslot\,\cite{Heslot1985}, used in essential ways in H.-T. Elze\,\cite{Elze2014},  the complex-valued wave functions arise as \(\psi=x+ip\), where \(x\) and \(p\) are integer valued conjugated "coordinate" and "momentum" variables for linear cellular automata. }

%%%%%%%

\newsecl{Massless particle in a box}{box}
The harmonic oscillator is closely related to a massless relativistic particle on a lattice (lattice length \(\d x\)\!) inside a box with length \(R=\ell\,\d x\), with hard walls at the edges:
\be H=|p|\equiv\s\, p\ ,\qquad x=\tl k\,\d x\ ,\quad \tl k \in[\,0,\, \ell\,]\ . \qquad  \tl k \ \hbox {integer.}\eel{boxham}
Here, the  ontological variables  are \( \tl k\), and \(\s=\pm 1\); they are related to the variable \(k\) in eq.~\eqn{ontok}, except that we take the particle to bounce to and fro:
\be  k(\tl k,\s)=\ell+\s(\tl k -\ell)\ ,\qquad \tl k\in[\,0,\,\ell\,]\ ,\quad k\in[\,0,\,2\ell\,]\ . \eel{kktilde}
This says that \(0\le \tl k <\ell\), while the velocity \(\pa H/\pa p\) flips when \(\tl k\) reaches a wall.

We see that now, in the \(\ell\ra\infty\) limit, this becomes a model for the free, relativistic massless particle on an infinitely fine one-dimensional lattice, with walls at its edges. The
velocity is fixed apart from its sign \(\s\). Keep in mind that the position operator \(x\) and the momentum operator \(p\) used here differ from \(x\) and  \(p\) that we used for the oscillator, eq.~\eqn{xpellcomm}, which is why now the Hamiltonian looks different. This section is merely to point out that one system can be transformed into the other. A possible advantage of the description in this section is that, here, \(x\) itself is ontological.

Note that the energy spectrum of a relativistic massless particle in the box is linear in the momentum \(p\), and the eigenvalues of \(p\) in the box are equidistant, and this is why this system can be mapped easily onto the harmonic oscillator.\fn{An apparent degeneracy of the Hamiltonian \eqn{boxham} with \(\s\) can be lifted by carefully imposing the boundary conditions needed for realising the reflections at the edges. We did not
include these here in order to avoid inessential complications, but the reader can derive them by unfolding the box where \(\tl k\) is living, to become the periodic box for \(k\) that has twice that length, see Eqs.~\eqn{kktilde}.} See also the last paragraph of this paper.

\newsecl{Momentum dependent kinetic term}{kin}

As already stated in Section~\ref{continuous}, one might wish to find an ontological interpretation of systems having a Hamiltonian of the form 
\be H=T(\vec p\,)+V(\vec x\,)+\vec A(\vec x\,)\cdot\vec p\ .\eel{HTVA} 
In the general case, neither \(x\) nor \(p\) can be considered to be ontological, since they both evolve as superpositions. However, in some special cases, a variable \(y(t)\) can be found that is ontological. The general rule is that we should search for operators such that they commute with themselves at all times, and  also with the commutator of the Hamiltonian and these observables (that is, their time derivatives).

We now consider the Hamiltonian \eqn{HTVA}  in the continuum case, in one dimension (so that we omit the vector symbols) and with \(V(x)=0\).  In this case, the vector potential field \(A(x)\) can be gauge transformed away, therefore we also put \(A(x)=0\). Define accolades to symmetrise an expression:
\be \{A\ B\}\equiv \half (AB+BA)\ . \eel{accdef}
Then we define the operator \(y(t)\) as
\be y\equiv\{x\,\frac 1{v(p)}	\}\ ,\qquad v(p)=\dd T(p)/\dd p\ .\eel{yx0}
It can be inverted:
 \be x=\{y\ v(p)\}\ . \eel{xy0}
 One easily derives that, indeed, 
 \be \frac{\dd}{\dd t}y(t)=1\ . \ee
 
Demanding both the operator \(v(p)\) and its inverse \(v^{-1}(p)\) to be sufficiently regular and unambiguous, forces us to keep the sign of \(v(p)\) constant.
  
 Note that, in the case where \(T(p)\) is quadratic in the momentum  \(p\),  \(v(p)\) is proportional to the square root of the energy; here again, we observe this square root relating the ontological variable with the more familiar \(x\) coordinate that we saw in Section~\ref{SU2}.
  
It is an interesting exercise to compute the inner product between the eigenstates of this ontological parameter \(y\) and those of the conventional \(x\) operator:
\be \bra x|y\ket=\int\dd p\bra x|p\ket\bra p|y\ket=\frac 1{2\pi}\int\dd p\sqrt{v(p)}\,e^{i(xp-yT(p))}\ . \eel{xybraket}
 It reduces to a Dirac delta function as soon as \(T(p)\) is linear in \(p\). In the more general case, unitarity demands that the function \(T(p)\) can be inverted, since only then the Cauchy integrals needed to prove unitarity of eq.~\eqn{xybraket} close. If \(T(p)\) can only be inverted on a finite interval of the values for \(T\) (or as the half-line rather than all values on the real line), then \(y\) is restricted to the values dual to that set.
 
 We decided not to pursue this analysis further since there seems to be a more imminent problem: the operator \(y(t)\), as defined in Eq.~\eqn{yx0}, in general, seems to be quite non-local. This is why the ontological variables derived here will not be used to replace space-like coordinates, but rather field-like variables, as in second-quantised field theories, which are explored in Section~\ref{tobe} and \ref{interact}.
 
 Of course we would also like to understand systems that do have effective potential functions \(V(x)\), and they should apply to higher dimensions as well. We shall home in to the completely general Hamiltonian in these last two sections, where we shall see that extra, high energy degrees of freedom will be needed in general.

\newsecl{Beables, changeables and superimposables.}{tobe}
What we call ontological, or deterministic, quantum mechanics is a particularly interesting subset of quantum systems where Hilbert space can be set up in terms of operators we call \emph{beables}, a phrase that was introduced by J.S.~Bell\,\cite{Bell-1964}\cite{Bell-1982}. 
These are (a set of) operators that all commute with one another at all times, so that, if we have a coordinate frame where at time \(t=0\) all beables are diagonalised, they continue to be diagonalised at all times,\fn{A special piece of insight is that \emph{measurement devices} will also be diagonalised, so that a measurement will always give unique, `beable' answers. In constructing models for experimental set-ups, one is free to choose the initial state as any wave function one likes, but if the beables are not all diagonalised, the final result will also come as a superposition. 
However if we postulate that the universe started in an eigenstate of all beables, the final measurement will also be unique. This can be used as a natural explanation for the `collapse of the wave function'.} and consequently, the evolution is completely classical. In this basis, the evolution operator \(U(t)=e^{-iH t}\), at distinct times \(t=t_i\), is a matrix containing only ones and zeros, see Eq.~\eqn{ontok}. We then refer to operators \(H\) and \(U(t)\) as \emph{changeables}: they act on the eigen states of the beables merely by replacing these eigen states with other eigen states. Both beables and changeables form small subsets of all possible operators. The generic operators are superpositions of different possible beables and changeables, and so we refer to these remaining operators as \emph{superimposables}.

In Section \ref{SU2}, the beables are the operators \(k\), and their eigen states are \(|k\ket^\ontt_t\). The operators \(U(\d t)\) (where \(\d t\) may have to be chosen to form a time-like lattice) and the associated Hamiltonian \(H\), or \(L_3\), are changeables, while operators such as \(L_1\) and \(L_2\) are superimposables.

In Section \ref{box}, \(\tl k,\ \s\) and \(k\) are beables, while \(H\) and \(p\) are changeables. In Section \ref{kin}, the operator \(y\) is the beable. \(T(p)\) is the Hamiltonian, and as such may serve as a changeable. The original position operator \(x\)  is merely a superimposable. 

Consider in particular the harmonic oscillator. Its mathematics is exactly as in Section \ref{SU2}, if now we take the limit \(\d t\ra 0\). The equations for the harmonic oscillator are written in terms of the familiar annihilation operator \(a\) and creation operator \(a^\dag\):
\be H=\half\w(p^2+x^2-1)=\w\,a^\dag a\ ,\quad & a=\fract 1{\sqrt 2}(p-ix)\ ,\qquad a^\dag=\fract 1{\sqrt 2}(p+ix)\ , &\nm\\[3pt]
	& p=\frac 1{\sqrt 2}(a+a^\dag)\ ,\qquad x=\frac i{\sqrt 2}(a-a^\dag)\ .&\nm \\[3pt]
	& [\,x,\,p\,]=i\ ,\qquad [\,a,\,a^\dag]=1\ .&    \eel{harmham}

In conventional quantum mechanics, all known operators are superimposables, except possibly for the Hamiltonian, which could be a changeable, if we would have been able to identify the complete set of beables. According to the \emph{cellular automaton theory of quantum mechanics}\,\cite{GtHCA}--\cite{BJV-2000}, the complete Hamiltonian of all physics in our universe happens to be a changeable. This theory can only be verified if we can also identify the ontological variables, the beables, and in practice this is hard. This paper is an attempt to pave the way to such a description of our universe.

In the next Section, we seek to describe the beables and changeables in terms of the  operators such as \(x,\ p,\) and \(H\) of finite, periodic cells. We are now in the limit \(N\ra\infty\), which means that we have \(\ell\ra\infty\), so that the `angular momenta' of Section~\ref{SU2} are almost classical. In practice, one frequently needs to switch between the strictly continuous case\fn{From a physical point of view, the distinction between continuous and step-wise evolution laws is less significant than one might think. One may or may not be interested in what happens between two distinct time steps, while what really matters is what happens after longer amounts of time.} and the case with finite time steps \(\d t\).

In the continuous case, the easiest changeable operator is  \(H=\w a^\dag a=\half \w(p^2+x^2-1)\). When \(\d t\) is taken to be finite, one needs the evolution operator \(U(\d t)\) to describe the motion of a beable \(k\) forward by one step, see Section~\ref{SU2}:
\be U(\d t)|k\ket^\ontt=e^{-iH\d t}|k\ket^\ontt = |k+1\!\mod N\,\ket^\ontt\ . \ee
We can also say that the ontological angle \(\vv\) in Section \ref{wave} is rotated by an angle \(2\pi/N\). Its angular rotation frequency is \(\omega\). This rotation is also generated by \(L_3\). 

What is the angle \(\vv\) in terms of the standard harmonic oscillator operators? From Eq.~\eqn{FF} in Section \ref{SU2}, we see that
\be e^{i\vv}|n\ket^E =|n+1\ket^E\ ,
\ee
and since \(a^\dag |n\ket^E=\sqrt{\fract H{\w}}\,|n+1\ket^E\), we find\fn{There is no problem here with the zero eigenvalues of \(H\), since in both equations \eqn{ontophi}, the number inverted is always 1 or bigger.}
\be e^{i\vv}=(\fract H{\w})^{-\halff}\,a^\dag\ ,\qquad  e^{-i\vv}=(\fract H{\w}+1)^{-\halff}\,a\ .\eel{ontophi}
\(\vv\) is the beable of the harmonic oscillator. By taking powers of the operators \(e^{\pm i\vv}\), we get \(\cos\k\vv\) and \(\sin\k\vv\) that, together, contain all desirable information concerning the ontological state the oscillator is in.
It also applies to the relativistic particle in a box, section~\ref{box}.

Next, let us construct a complete list of all changeables. One changeable is easy to recognise: the Hamiltonian itself. However, in the next section,  we set up the procedures to obtain all possible interaction Hamiltonians, and for that, we need different changeables. Consider the basis of all ontological states, called \(|k\ket^\ontt\) in section \ref{SU2}. Then a generic changeable operator interchanges two such elements, say \(|k_1\ket\) and \(|k_2\ket\). We write it as\,\fn{We often suppress signs and phase factors, when these have no effect on what happens physically in the ontological system. In general, adding phase factors will not help us to describe more general physical systems; at a later stage, however, phase factors can serve some important mathematical purpose.\\ A reader might wonder how \emph{complex numbers arise in the Hamiltonian?} This `mystery' disappears when complex numbers are treated as pairs of real numbers. In practice, complex amplitudes are often linked to the conservation of matter, or more precisely: baryon number.}
\be G_{12}=|k_1\ket\,\bra k_2|+|k_2\ket\,\bra k_1|\ . \eel{interchange}

The combination of the two terms in Eq.~\eqn{interchange} will be needed to preserve hermiticity, as will be clear in the next section. \(G_{12}\) is just one possible changeable. If we combine it with all other expressions of the same form, \(G_{ij}\), we may obtain the complete set of all unitary permutations. 

One further generalisation is conceivable: if we have \emph{many} independent harmonic oscillators (or periodic subsystems), we get a more generalised system with an ontological evolution law. 
The interchange operators \( G_{ij}\) must then be allowed to interchange the states of different oscillators.

In practice, we shall need to consider in particular an operator that only interchanges two states in one given periodic cell / harmonic oscillator, but only if the beables in a neighbouring cell -- or beables in several neighbouring cells -- take one particular value. This set is not completely but almost general, so that we can now perform our next step: construct non-trivial interaction Hamiltonians.

The harmonic oscillator Hamiltonian \(H_0\) itself may be regarded as the infinitesimal interchange operator for all pairs of neighbouring\,\fn{to describe the continuum limit, we need to scale the variable \(k\) so that the steps \(\d k\) are no longer 1, as before, but infinitesimal.} states \(|k_1\ket^\ontt\) and \(|k_1+\d k\ket^\ontt\). If we let its amplitude depend on where we are in \(k\) space, we have a small but very special class of changeables that we shall also need.

\newsecl{Ontological interactions}{interact}
So-far, our models were of an elementary simplicity. Now, we can put them together to obtain physically more interesting systems. Let us start by having a large class of small, independent, ontological ``cells", listed by an index \(i\). All of them are so small that, as soon as the influences of other cells are shut off, their own internal motion forces them to be periodic. 

In  more advanced future approaches, one might wish to avoid these elementary cells, but for our present purpose they appear to be quite useful.

Let the time step \(\d t\) be 1, and in each cell \(i\) we have a variable \(k^{(i)}\), with an associated momentum operator \(p^{(i )}\).  \ A  Hamiltonian \(H_0\) forces all data \(k^{(i)}\) to make one step forward at every time step, with periodicity \(N^{(i)}\)\,(allowed to be different for all \(i\)):
\be H_0=\sum_i p_i \ ,\qquad |k^{(i)}\ket_{t+1}^\ontt=|k^{(i)}+1\ket_t^\ontt\ , \qquad |N^{(i)}\ket^\ontt=|0^{(i)}\ket^\ontt\ . \ee
The basis elements of the combined states are written as
\be |\vec k\ket_t^\ontt=\prod_{i}|k^{(i)}\ket^\ontt_t\ . \ee

The ontological interaction to be considered next is an extra term to be added to \(H_0\) such  that the \(k^{(i)}\) evolve in a more complicated way. First, consider just one cell, \(i=1\), and consider two special values, \(k_1\) and \(k_2\), with 
 \(k_2>k_1\). We now ask for an extra term such that, as soon as the value \(k=k_1\) or \(k_2\) is reached, it switches to the other value \(k=k_2\) or \(k_1\) (by adding the difference, \(\D k=\pm(k_2-k_1)\) to the value of \(k\)). In addition to this exchange process, we keep the term \(H_0\)  in the Hamiltonian, which ensures that at all times also \(k\ra k+1\) (there is a danger with the non-commutativity of the two terms, but in the continuum limit this ambiguity disappears, while in the discrete case one must ensure unitarity by demanding that every state \(|k\ket\) evolves into exactly one other ket state \(|k'\ket\)).
  
At first sight, this gives a rather trivial effect: either the stretch \([k_1 , \,k_2]\) is skipped, or the system moves within the stretch \([k_1,\,k_2]\) forever. This means that, physically, we just changed the period of the motion. Nevertheless this is an important interaction, as we shall see. We write the Hamiltonian as
\be H=H_0\pm\pi|\psi\ket\,\bra\psi|\ ,\qquad |\psi\ket=\fract1{\sqrt 2} \big(|k_1\ket-|k_2\ket \big)\  ;\eel{fundH}
both signs are allowed. 
One may derive that\fn{The situation becomes more delicate in case more of such terms are added to the Hamiltonian while \(\d t=1\). To do this right, it will be advisable to keep the continuum limit \(\d t\ra 0\) and ensure that the exchanges are kept at a well-defined order. On the other hand we can guarantee that for any ontological (classical) interaction, a well-defined quantum Hamiltonian exists.}, indeed, the evolution operator \(U_{\d t=1}\) contains an 
extra factor \(e^{-\pi i}=-1\) whenever the (normalised) state \(|\psi\ket\) is encountered. The state \(|\phi\ket=\fract1{\sqrt 2}\big(|k_1\ket+|k_2\ket\big)\) is orthogonal to \(|\psi\ket\) and is therefore not effected. One sees immediately that the net effect of \(U_{\d t=1}\) is that all \(k\) values made one step forward, while \(|k_1\ket \leftrightarrow |k_2\ket\)\,.

Thus we introduce the interaction term \(H^\intt=\pi|\psi\ket\,\bra\psi|\) as an ontological interaction. However, its effect is very strong (it only works with the factor \(\pi\) in front), whereas in general, in quantum physics, we encounter interaction terms of variable strengths, whose effects can be much more general. 

Therefore, we consider a further modification. Let us impose the condition that\\[5pt]
\emph{the interchange \(k_1^{(1)}\leftrightarrow k_2^{(1)}\) only happens if in a neighbouring cell, say cell \(\# 2\), \(k^{(2)}\) has a given value, say \(k^{(2)}=r\)}\,:
\be H^\intt=\pi\,|r^{(2)}\ket\,\bra r^{(2)}|\ |\psi^{(1)}\ket\,\bra\psi^{(1)}|\ . \eel{Hintt}
Intuitively, one might suppose that this interaction is much smaller, since in the majority of cases, when \(k^{(1)}\) takes the value \(k_1\) or \(k_2\), it must be rather unlikely that \(k^{(2)}\) equals \(r\). Does this argument make sense?

Yes, indeed it does, in the following way:\\[5pt]
In the real world, \(\d t\) will be extremely tiny, perhaps as small as the Planck time, some \(10^{-44}\) seconds. Suppose that we only know the interaction Hamiltonian of our `Standard Model' up to some maximum energy \(E^\mmax\) of all scattering processes. In practice,
\be E^\mmax\ll 1/\d t\ . \ee
If we limit ourselves to states composed of lower energies only, we can never make Gaussian peaks much sharper than
	\be\bra k^{(2)}|\psi^{(2)}\ket \approx \left(\fract \a{\pi}\right)^{1/4}e^{-\halfje\a(k^{(2)}-r)^2}\ ,\ee
where \(\a=\OO(E^\mmax)^{-2}\). This implies that the amplitude of the interaction Hamiltonian \eqn{Hintt} will be very tiny indeed. We can add many such terms to our Hamiltonian before the effects become sizeable.

We should not be concerned that integrating many of such interaction terms to obtain the elementary evolution operator \(U_{\d t}\) over one time step, might add terms of higher order in \(\d t\) that violate the condition that this term is ontological. Although such considerations may well be important for actual calculations, they do not affect the principle that we can generate a large class of interaction Hamiltonians along these lines.

Which Hamiltonians can we obtain now? the answer may come as a surprise: \\[5pt]
\emph{All Hamiltonians acting in elementary cells, and generating interactions between neighbouring cells, can be obtained from ontological interactions along these lines!} They certainly do not need to commute.

Suppose we have been adding ontological interactions of the type described above, but we still need one matrix element \(\bra x_1|H|x_2\ket\), where \(|x_1\ket\) and \(x_2\ket\) are elements of any basis one may wish to employ. Then all we have to do is take the set of ontological states \(|\vec k\ket\) that we started from. Apply the unitary matrix \(\bra x|\vec k\ket\) that links the basis of ontological states \(|k\ket^\ontt\) to the states \(|x\ket\), to rewrite the desired Hamiltonian in the ontological basis. Consider any of its off diagonal terms \(\bra k^{(1)}_1|H^\intt|k^{(1)}_2\ket\). Then, according to Eq.~\eqn{Hintt}, the missing term can be reproduced by an ontological exchange contribution of these two states in this particular cell. 

As for the Standard Model itself, we know that its Hamiltonian does not connect cells far separated from one another, because the Hamiltonian density of all quantum field theories are known to be local in the sense of what is called `no Bell telephone' among philosophers:

 The Hamiltonian density \(\HH(\vec x\,)\) in 3-space \(\vec x\), is such that, at different locations \(\vec x\), these Hamiltonian density functions commute:
\be [\HH(\vec x\,),\,\HH(\vec x\,'\,)]=0\quad\hbox{if}\quad \vec x\ne\vec x\,'\ .\eel{lightcone}
If the speed of signals is bounded, this leads to commutation of operators outside the light cone.

J.S.~Bell himself\,\cite{Bell-1982}
 called this `local commutativity', but insisted that tighter definitions of causality -- forward and backwards in time -- are needed if one wants to compare quantum mechanics with deterministic models. He was criticised
on this point\,\cite{Brans-1987}\cite{Vervoort-2013}, and also the present author disagrees;  here we just remark that further equations that would be tighter than Eq.~\eqn{lightcone} are not needed. We derive from that equation that  off-diagonal matrix elements of the Hamiltonian vanish when they refer to states in cells that are separated far from one another (far meaning far in units of te Planck length!). We connect this with our interaction equation \eqn{Hintt} to conclude that exchange interactions between ontological states that are far separated form one another are not deeded. Our models should violate Bell's theorem and the inequalities arrived at by Clauser et al\,\cite{CHSH-1969}
simply because our interactions appear to generate quantum field theories.

We do note that, besides generating one off-diagonal term of the interaction Hamiltonian matrix, and its Hermitian conjugate, the interaction Hamiltonian \eqn{Hintt} also affects the diagonal terms at \(k^{(1)}_1\) and \(k^{(1)}_2\). therefore, we might have to readjust all diagonal terms of the Hamiltonian. In the continuum limit, \(\d t\ra 0\), this is easy. It just means that the speed at which a given ontological state may make a transition to the next state, may have to be modified.

We also obtained a bonus: the ontological theory is local up to Planckian distance scales, as soon as the commutator rule \eqn{lightcone} is obeyed by the quantum system that we wish to reproduce in ontological terms.

We find that by including ontological exchange interactions between the ontological states, and by adjusting the speed of the evolution, we can create a quite generic quantum mechanical Hamiltonian. The model we get is a \emph{cellular automaton}\,\cite{Ulam-1952}--\cite{Wolfram-2002}, exactly as we described in Ref.\,\cite{GtHCA}, but now we find the systematic prescription needed to let this automaton act as any given local quantum field theory.

It seems now that almost any basic interaction can be obtained, but there are important questions that we have not been able to answer:

\emph{We started with cells having no mutual interaction at all. In terms of elementary particle theories, this means that we start with infinitely heavy elementary particles. Then kinetic terms can be added that should lead to finite mass particles. What is the most efficient way from here to realistic quantum particle theories?} 

Our difficulty is that we do not quite understand how to construct ontological free field theories, describing light or even massless particles. We need a theory starting from ontological \emph{harmonically coupled} oscillators that obey some form of locality. This seems to be a purely technical problem now, which ought to be resolved.

Another question: \\
\emph{The theory for the interactions between elementary particles features quite a few continuous symmetry properties. How can we reproduce such symmetries?} Since our cells tend to be discrete, it is hard to impose continuous symmetries, in particular Lorentz invariance. On the other hand, it is generally expected that most if not all \emph{global} continuous symmetries cannot be exactly valid.

\newsecl{How a sieve can connect classical theories with quantum mechanical ones.}{connect}

Note that our sieve mechanism can easily be generalised to do the following:\\[5pt]  \emph{
Given two different ontological theories, with Hamiltonians \(H_1\) and \(H_2\), which do not have to commute. Then, adding a fast `sieve' variable enables us to describe a system that, at sufficiently low energies, is described by a Hamiltonian}
\be H_3=\a H_1+\b H_2\ , \ee
where indeed quantum effects might lead to quantum interference, while, at the microscopic level, the theory is still deterministic. 

Such a result seems almost too good to be true. Can we really re-write quantum mechanical systems in terms of deterministic ones? \emph{How does it go in practice?} \,Which calculations turn a classical system, with exchange interactions and sieves or whatever else, into quantum mechanics? We here briefly analyse how to address this question.

We consider the simplest model of a sieve, giving rise to a Hamiltonian that generates superpositions as soon as the sieve is turned on, and then compare the quantum system with the classical one. The most important feature is that we have two (time) scales.

Our model is sketched in Fig.~\ref{sieve.fig}. Its states are described as \(|x\ket\) where the fundamental, ontological variable  \(x\) sits in a box with size \(L\) and periodic boundary conditions, \(|x+L\ket=|x\ket\).

 \(x\) starts out being a beable. The unperturbed Hamiltonian is \(H_0=p_x\), so the velocity is 1, the system circles around in its box. Now, we wish to turn it into a real quantum variable as in the previous section. We add the term \(\alpha\pi|\psi\ket\,\bra\psi|\), with \(|\psi\ket=\fract 1{\sqrt 2}\big(|0\ket-|A\ket\big)\) to the Hamiltonian, that had the value \(H_0=p_x\) in the unperturbed case. If \(\a=0\) we have the original model. If \(\a=1\), the Hamiltonian generates the exchange
 \(|0\ket\leftrightarrow |A\ket\) in \(x\)-space. In that case, the \(x\) variable either stays inside the region \([0,\,A]\) or inside the region \([A,\,L]\) This makes it also a beable.

\begin{figure}[h]\bcenter\widthfig{350pt}{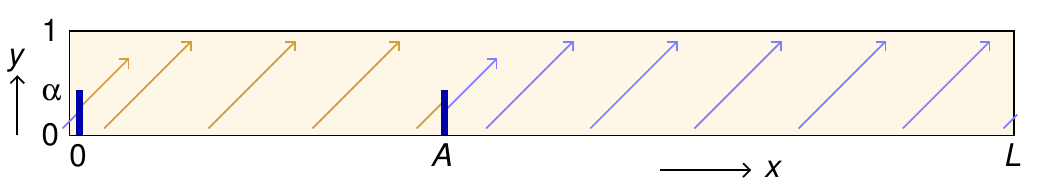}\ecenter\caption{\footnotesize The variables \(x\) (horizontal, period 
\(L\gg 1\)) and the sieve variable \(y\) (vertical, period \(1\)), together moving in the direction of the arrows. See text.}\labell{sieve.fig}\end{figure}

However, if \(\a\) has any other value, the Hamiltonian contains a non-diagonal element, together with its Hermitian conjugate, resulting in quantum interference. According to Section~\ref{interact}, we can mimic the suppression factor \(\a\) by adding a fast variable \(y\), such that only in the fraction \(\a\) of all states \(y\) can be in,  the exchange takes place. This we describe by replacing
\be \a=\sum|r\ket\,\bra r|\ \eel{rproject}
in \(y\) space. The period of \(y\) is here taken to be 1, much shorter than the period \(L\) for \(x\).

Classically, if the variable \(y\) takes one of the values \(r\) of the projection operator \eqn{rproject}, precisely at the moment when \(x=0\) or at the moment when \(x=A\), the exchange \(|0\ket\leftrightarrow |A\ket\) takes place, otherwise it does not.
 
We argued in Section~\ref{interact} that, as long as the variable \(y\) stays in its lowest energy state, the equations of motion for the quantum system and the classical system are the same. Now, we are in a position to check this.

The unperturbed Hamiltonian is now \(H_0=p_x+p_y\); \(x\) and \(y\) move at the same speed \(v=1\), but \(y\) makes many oscillations during the time \(x\) makes one oscillation. Thus, the system moves in the direction of the arrows in Fig.~\ref{sieve.fig}.

We claim that, in quantum language, the \(x\) variable obeys its effective Hamiltonian equations, i.e. the Schr\"odinger equation -- including superposition effects (when \(\a\ne 0\) and \(\ne 1\). What happens classically?

The system evolves along the diagonal arrows in Fig.~\ref{sieve.fig}. The boundary conditions are: \(x\) is periodic with period \(L\) and \(y\) is periodic with period 1. Then, we have the `sieve', consisting of two partial walls of length \(\a\) in the \(y\) direction, one at \(x=0\) and one at \(x=A\). The rule is that, if \(x=0\) or \(x=A\) while \(0<y<\a\) then the states \(|0,y\ket\) and \(|A,y\ket\) are interchanged, after which the evolution continues in the direction of the arrows. We see that, classically, an infinite orbit results. If the lengths \(A\) and \(L\) have an irrational ratio, the orbit never closes. If we write this in terms of a wave function \(\bra x,y|\psi(t)\ket\), we get some sort of fractal.

The same orbit can now be described in quantum notation.  In the bulk region, we have a wave function, which we write as  
\be \psi(x,y,t)=\psi(x-t,y-t,0)=\sum_{n=0}^\infty\psi_n(x,t)\,e^{2\pi i n y-2\pi i n t}\ . \eel{psin}
We must restrict ourselves to non-negative \(n\), as should be clear from the earlier sections of this paper: there is a lowest energy mode, which was tuned to the value \(n=0\) (a procedure that we can also apply to the \(x\) coordinate, but for the time being, we keep the \(x\)-space notation as is).

Now here is how we can see the effect of the sieve. We consider the waves \(\psi^{0-}\) near the point \(x=0\) entering at \(x<0\) and the waves \(\psi^{A-}\)  near \(x=A\), entering at \(x<A\). At \(x>0\) the waves \(\psi^{0+}\) are leaving, and at \(x>A\) the waves  \(\psi^{A+}\) are leaving. Write 
\be \psi^{1\pm}=\fract 1{\sqrt 2}(\psi^{0\pm}+\psi^{A\pm})\ , \quad\hbox{and}\quad 
\psi^{2\pm}=\fract 1{\sqrt 2}(\psi^{0\pm} -\psi^{A\pm})\ . \ee
Then, at the sieve, 
\be &\psi^{1+}=\psi^{1-} \quad\hbox{for all} \ y\ ,&\nm\\
&\psi^{2+}=\psi^{2-}\quad\hbox{if} \quad \a<y<1\ ;\qquad \psi^{2+}=-\psi^{2-}\quad\hbox{if} \quad 0<y<\a\ . &\ee

This is now easy to rephrase in terms of properties of the functions \(\psi_n(x,t)\). We also split these into functions \(\psi_{n}^{1\pm}\) and \(\psi_{n}^{2\pm}\). Since
\be \psi_n(x)&=&\int_0^1\dd y\,e^{-2\pi in y}\,\psi(x,y)\ ,\nm\\[3pt] \hbox{and}\quad\int_0^1\dd y\, e^{2\pi i\ell y}\,\sign(y-\a)&=&\frac{-i}{\pi\ell}\big(1-e^{2\pi i\ell\a}\big)\quad\hbox{if}\quad \ell\ \hbox {integer}\ne 0\ ,\nm\\
&=&1-2\a\quad \hbox{if}\quad \ell=0\ ,\ee 
we derive
\be \psi_{n}^{1+}&=&\psi_{n}^{1-}\ ;\nm\\
 \psi_{n}^{2+}&=&(1-2\a)\psi_{n}^{2-}+\sum_{\ell\ne 0}\frac i{2\pi\ell}\big(1-e^{2\pi i\ell\a}\big)\psi_{n+\ell}^{2-}\ ,\ee
so that we find
\be \psi^{0+}_{n}&=&\psi^{0-}_{n}(1-\a)+\a\,\psi^{A-}_{n}+\sum_{\ell\ne n}\frac i{4\pi\ell}(1-e^{2\pi i\ell\a})(\psi^{0-}_{n+\ell}-\psi^{A-}_{n+\ell})\ ,\nm\\
\psi^{A+}_{n}&=&\psi^{A-}_{n}(1-\a)+\a\,\psi^{0-}_{n}+\sum_{\ell\ne n}\frac i{4\pi\ell}(1-e^{2\pi i\ell\a})(\psi^{A-}_{n+\ell}-\psi^{0-}_{n+\ell})\ .\eel{result}

We see that the first terms contain the effect of a quantum superposition. Only when \(\a=0\) or 1, this represents classical motion in \(x\)-space.  For the other values of \(\a\), Eq.~\eqn{result} looks quantum mechanical. Now we know that, provided we include the extra terms, this is classical motion after all. But the extra terms only contain the high energy states 
\(\psi_n\). We see in Eq.~\eqn{psin} that the \(n>0\) modes carry a large amount of energy, so that ignoring them does not affect much the physical nature of these transitions.

\newsecl{Conclusions}{conc}

We conclude that an ontological theory for the basic interactions is quite conceivable. We find that one can postulate the existence of `cells' that each contain one or more variables; these variables are postulated each to move in periodic orbits (circles) as long as other interactions are absent. Then, we carefully postulate two kinds of interactions, together generating behaviour that is sufficiently general to mimic any fundamentally quantum structure. One fundamental force is generated when a variable in one particular cell changes position with a variable in either the same cell or in its immediate vicinity. This exchange interaction will be associated with non-diagonal terms in what later will be our Hamiltonian. To weaken the force, and to make it fundamentally quantum mechanical, we need a sieve, in one or more cells, again in the immediate vicinity (or in the same cell). The ontological variable(s) of the sieve cell will only allow for the given exchange process if the sieve variable takes some pre-assigned value(s). Since we can choose which variables to exchange, as well as the strength of the sieve, this process will generate almost any Hamiltonian.

To then adjust the diagonal terms of the Hamiltonian, all that needs to be done is to adjust the velocities of the variables, depending on their positions in the cells. All taken together, we have as many degrees of freedom to adjust, as there are terms in our (Hermitian) Hamiltonian. This gives us confidence that our procedure will work, regardless the quantum model we are attempting to `explain' in ontological terms.

Actually, we still have a lot of freedom: we can derive which exchange interactions between elementary, ontological states will be needed in order to obtain agreement with today's Standard Model, but the details of the sieve, being the constraints imposed by cells neighbouring a given cell, as the ones of the cell \# 2 described in Eq.~\eqn{Hintt}, will be difficult to derive or even guess.
It is clear however, that the effects of the sieves will be almost continuously adjustable, depending on the maximal energies that we allow for our particle-like degrees of freedom (low energies will imply that most sieves are shut off, so that the effects of quantum forces get weak at energies very low compared to the elementary scales of our cells). 

Mathematical tools of the kind presented in this paper will be useful to study the constraints imposed on any `unified field theory' by the condition that, at some special time- and distance scale, our world is controlled by ontological forces.

Hopefully, our general strategy (as published during a few decades by now) is becoming more transparent with these demonstrations. Our `cells' are labelled by an index \(i\), and we set up models for particles by arranging such cells in a lattice, called a `cellular automaton'. The ontological degrees of freedom, \(k^{(i)}\) are the positions of these points on their orbits. What has been achieved in this paper is that we identified a way to characterise generic ontological interactions between the cells, using the \emph{language} of quantum mechanics. The generic interactions, of which one may add a large number at each lattice point, take the following form:
\be H=H_0+H^\intt\ ,\qquad H_0=\sum_{\mathrm{cells}\ i} \,p^{(i)}\ , \eel{fundH0}
where \([k^{(i)},\,p^{(j)}]=i\d_{ij}\)\,, and
\be &H^\intt=\pi\sum|r\ket\,\bra r|\ |\psi^{(i)}\ket\,\bra \psi^{(i)}|\ , &\\
&|\psi^{(i)}\ket = \fract 1{\sqrt 2}\big(|k^{(i)}_1\ket -|k^{(i)}_2\ket\big)\ . \eel{fundH1}
Here, the notation used was the one describing discrete states. These must be replaced by the continuous states when the \(\d t\ra 0\) limit is taken. Hence, for now,  the coefficient \(\pi\) has to be 3.14\dots, since only with this unique strength, the system exchanges position \(|k^{(i)}_1\ket\) with position \(|k^{(i)}_2\ket\) in a deterministic manner. To make the interaction sufficiently small so that it can be inserted in a perturbative quantum field theory, we use one or more other state(s) \(|r\ket\) as projectors (calling them `sieves'). The strength of the interaction then reduces by factors \(\OO(\sqrt{E^\mmax})\) in units where the time step \(\d t\) is one. This is why we are led to consider the case where all particle energies are low compared to the energies associated to the very tiny time scale of \(\d t\). 

We saw in section~\ref{connect}, that any strength of the off diagonal terms of the Hamiltonian can be obtained by adding fast variables to the classical system, and although extra terms do arise, the fundamentally quantum nature of the interactions is not jeopardised. We find this result very important, it should be regarded as a strong indication that this is the way to interpret what goes on in our quantum world.

We have high energy modes, which are claimed not to ruin the quantum nature of the results of our calculation. This is also where thermodynamics may enter the picture: if we ignore high energy modes, those are exactly what is left of a fast moving auxiliary variable\,\fn{The question was asked by Hossenfelder and Palmer\,\cite{superdet-2020}.}.  Today, in practice, we have too little information to be able to make fundamental distinctions between quantum mechanical and classical behaviour: we only know the outcome of scattering experiments as long as the total energy is kept below the limit \(E^\mmax\). All our experiments are at temperatures too low to allow us to do the timing of dynamical variables sufficiently accurately. We are too close to just one physical state: the vacuum. Thus we must tolerate uncertainties in our descriptions. These are the quantum uncertainties.

Since quantum mechanical language was used throughout, and since the states \(|\psi^{(i)}\ket\) connect different basis elements, we see that non commuting interaction Hamiltonians emerge. The only important constraint on \(|\psi^{(i)}\ket\) is that it should connect only nearby ontological states, since only then the resulting Hamiltonian operator obeys locality, which is expressed uniquely in terms of commutators vanishing outside the light cone, Eq.~\eqn{lightcone}.

What remains to be done is to achieve more experience in constructing realistic models along these lines, check how  these models perform, and reach consensus about their usefulness.

The most important message, we believe, is that quantum mechanics should not be considered as mysterious, it is not fundamentally impossible to understand it from the perspective of classical logic, and the origin of the uncertainty relations can be understood. It all amounts to \emph{timing}, that is, if fast moving, space-like separated,  variables are involved in an interaction, it is fundamentally impossible to adjust their time variables sufficiently accurately.

The question whether time is discrete or continuous is physically unimportant; as soon as some description of a system clarifies it sufficiently well, there will be no need to split the time variable into even smaller segments. There is a practical difficulty:  only one variable needs to be handled as if continuous: the dynamics of the variable that is used as our clock. All other variables will be limited in number, as one may conclude from black hole physics: black holes can only come in a finite number of quantum states. This implies that, in contrast with our clock, the other variables may have to be kept on a discrete time lattice. So where does our clock come from? In our models, we can choose whatever pleases us, but to guess the right model may be not easy.

Some of our readers may find it difficult to believe that points running around on circles are equivalent to harmonic oscillators. Here, we would like to use an analogy with the science of planets. We can study distant planets through our telescopes, and detect many interesting properties, for which we can find equations. Yet there is one thing we shall never detect: their names. Similarly, we can never tell whether a harmonic oscillator is actually a point moving on a circle. 
The equations are equivalent. Everything else is name giving.

 \end{document}